\documentclass[10pt,twocolumn]{IEEEtran}

\usepackage{graphicx}
\usepackage{amsmath}
\usepackage{amssymb}
\usepackage[caption=false]{subfig}
\usepackage[noadjust]{cite}
\usepackage{float}
\usepackage{algorithm}
\usepackage{bibentry}
\usepackage{balance}
\usepackage{algorithm}
\usepackage{algorithmicx}
\usepackage{algpseudocode}
\usepackage{xcolor}
\usepackage{subfig}
\usepackage{wrapfig}

\graphicspath{{img/}}
\usepackage{url}
\begin{document}

\bstctlcite{IEEEexample:BSTcontrol}


\title{Base Station Antenna Uptilt Optimization for Cellular-Connected Drone Corridors
\thanks{This work is supported in part by the INL Laboratory Directed Research Development (LDRD) Program under DOE Idaho Operations Office Contract DEAC07-05ID14517.}\thanks{S. J. Maeng, M.M.U. Chowdhury, \.{I}. G\"{u}ven\c{c}, and H. Dai are with the Department of Electrical and Computer Engineering, North Carolina State University, Raleigh, NC 27606 USA (e-mail: smaeng@ncsu.edu; mchowdh@ncsu.edu; iguvenc@ncsu.edu; hdai@ncsu.edu).}\thanks{A. Bhuyan is with the INL Wireless Security Institue, Idaho National Laboratory, Idaho Falls, ID 83402 USA (e-mail: arupjyoti.bhuyan@inl.gov).}}

\author{\IEEEauthorblockN{Sung Joon Maeng, Md Moin Uddin Chowdhury, \.{I}smail G\"{u}ven\c{c}, Arupjyoti Bhuyan, and Huaiyu Dai}}%

\maketitle

\begin{abstract}
The concept of drone corridors is  recently getting more attention to enable connected, safe, and secure flight zones in the national airspace.  To support beyond visual line of sight (BVLOS) operations of aerial vehicles in a drone corridor, cellular base stations (BSs) serve as a convenient infrastructure, since such BSs are widely deployed to provide seamless wireless coverage. However, antennas in the existing cellular networks are down-tilted to optimally serve their ground users, which results in coverage holes if they are also used to serve drones. In this letter, we consider the use of additional uptilted antennas at cellular BSs and optimize the uptilt angle to minimize outage probability for a given drone corridor. Our numerical results show how the beamwidth and the maximum drone corridor height affect the optimal value of the antenna uptilt angle.
\end{abstract}

\begin{IEEEkeywords}
Antenna uptilt, drone corridor, national airspace (NAS), outage probability, UAV, UTM.
\end{IEEEkeywords}

\section{Introduction}
  
Drones, also known as an unmanned aerial vehicles (UAVs), are rapidly gaining attention due to a wide range of promising applications. Commonly referred drone applications include search and rescue for public safety, commercial delivery service, and surveillance,  among others~\cite{marojevic2020advanced}. According to the Federal Aviation Administration (FAA) forecast, 
the number of active UAVs is expected to reach 2 to 3 million by 2023~\cite{FAA1}. The framework for UAV traffic management  (UTM) in the airspace has recently been developed by the FAA and the National Aeronautics and Space Administration (NASA)~\cite{FAA2}. In this context, the concept of drone corridors has been gaining more attention, which serve as sky lanes that the UAVs are required to pass through for safe and secure 
flow of UAV traffic~\cite{bhuyan2021secure}.

In cellular wireless networks, studies of drone applications typically fall under two categories:  
1) base station mounted UAVs (UAV-BSs) that serve other ground and aerial users~\cite{xu2020multiuser,yapici2021physical}, and 2) cellular connected UAVs (C-UAVs) that are users from the perspective of ground BSs~\cite{miao2020location,d2020analysis}. The success of future C-UAV operations (and in some cases, UAV-BS operations) relies critically on beyond visual line of sight (BVLOS) connectivity in drone corridors~\cite{FAA2}.  For example, the United Nations Children’s Fund (UNICEF) launched a drone testing corridor that monitors natural disasters, provides Wi-Fi signals, and delivers medical supplies for humanitarian purposes \cite{UNICEF}. In addition, the drone taxi service operating in aerial corridors has been tested by several  companies. The optimal placement of the BSs and the number of antennas to support the drone corridor is studied in~\cite{singh2021placement}.

While cellular networks may provide reliable and seamless BVLOS connectivity to drones, the cellular BS antennas are optimized and down-tilted to best serve their ground users. As a result, this antenna setup is not appropriate to serve the drone corridor in the airspace. In this letter, we consider an additional set of uptilted cellular BS antennas to serve drones, and explore the optimal uptilt angles in order to provide a reliable coverage at a drone corridor.
As a first work in this research direction, we design the drone corridor in a 2D coordinate system and consider two adjacent BSs, positioned across the drone corridor, to take into account the effect of the interference signal from the neighboring BS. Extending the study to 3D drone corridor design will be considered in future work.
The contributions of our work can be summarized as follows: 1) We identify five unique cases for aerial coverage that are dependent on the uptilt angles and beamwidths of the ground BS antennas; 2) We derive closed-form expressions for the signal-to-interference plus noise ratio (SINR) outage probability and the average SINR over a drone corridor; 3) We find the optimal antenna uptilt angle that minimizes the SINR outage probability; 4) We show that the average SINR is maximized for an uptilt antenna angle in case~5 (see Fig.~\ref{fig:illu}); 5) We show the effect of the beamwidth and the maximum drone corridor height on the SINR outage probability.

\vspace{-0.5mm}

\section{System Model} \label{sec:system}\vspace{-0.5mm}

\begin{figure}[t]
    \centering
    \subfloat[Case 1]{\includegraphics[width=0.47\textwidth]{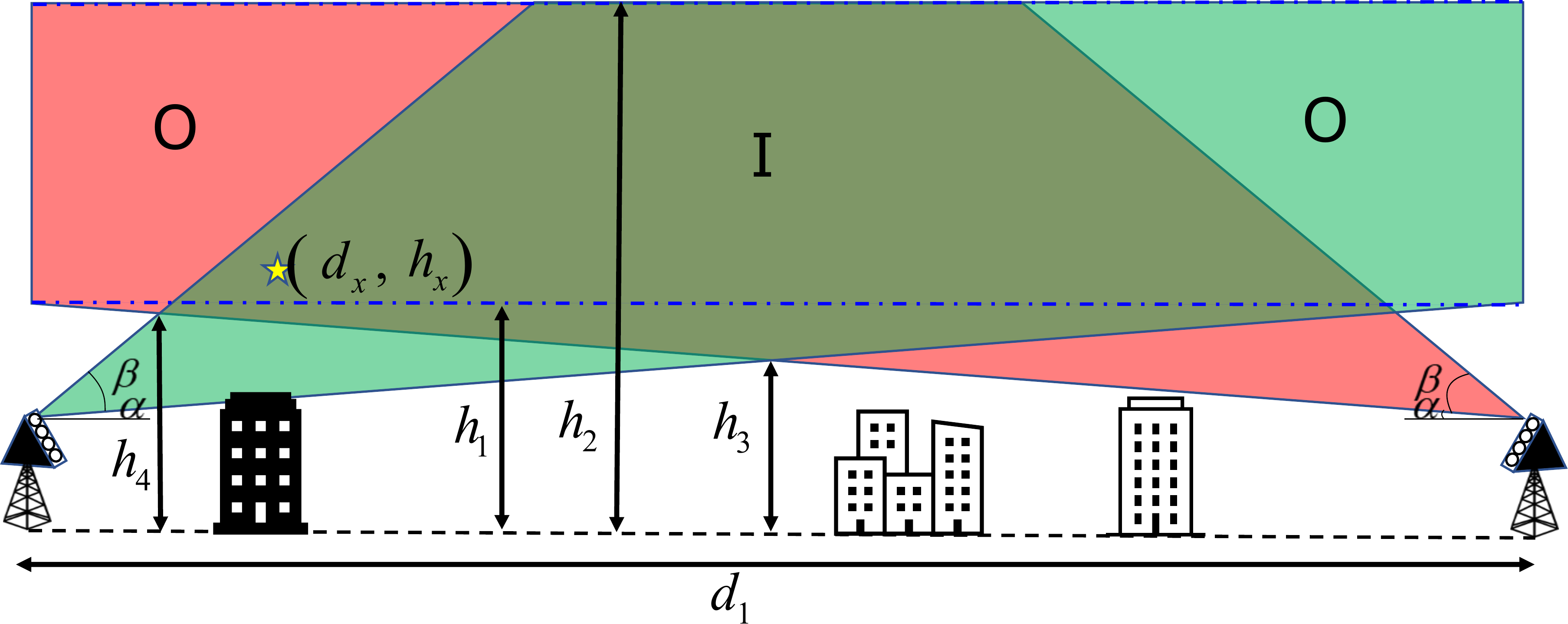}\label{fig:illu_c1}}\vspace{-1mm}
    
    \subfloat[Case 2]{\includegraphics[width=0.47\textwidth]{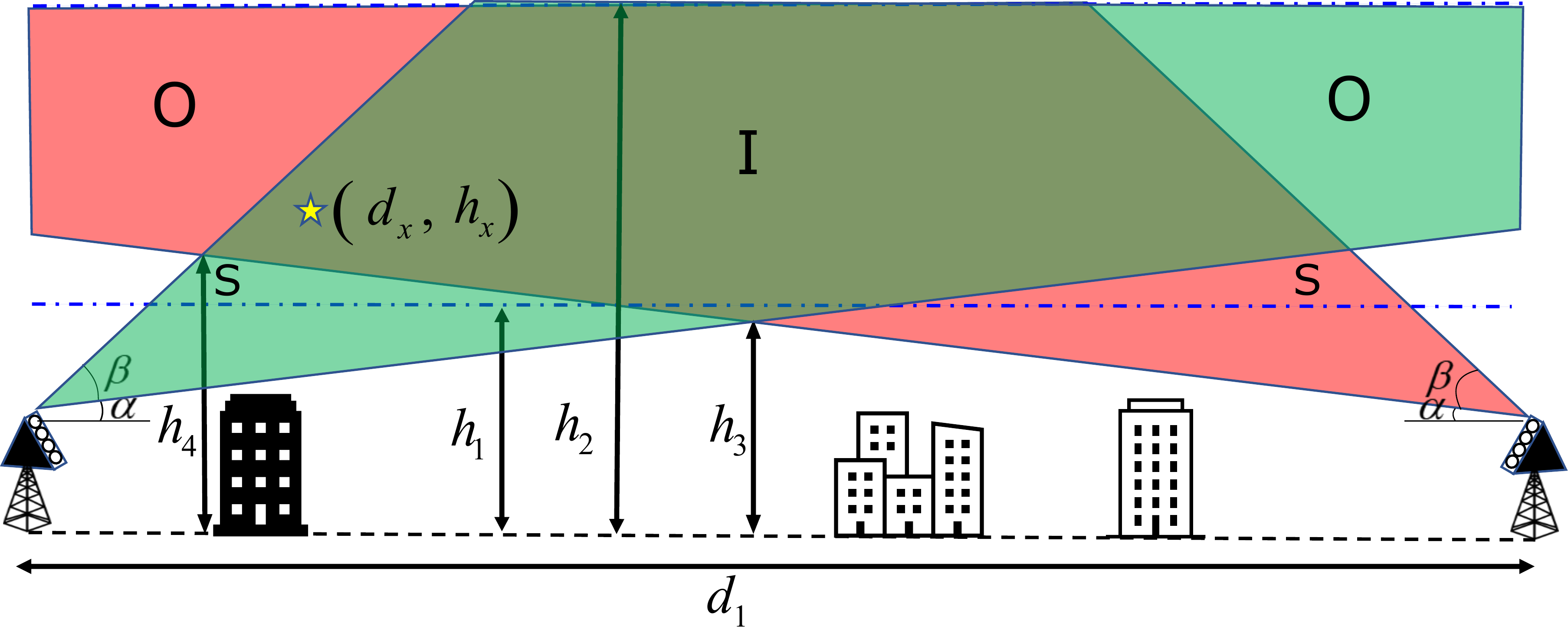}\label{fig:illu_c2}}\vspace{-1mm}
    
    \subfloat[Case 3]{\includegraphics[width=0.47\textwidth]{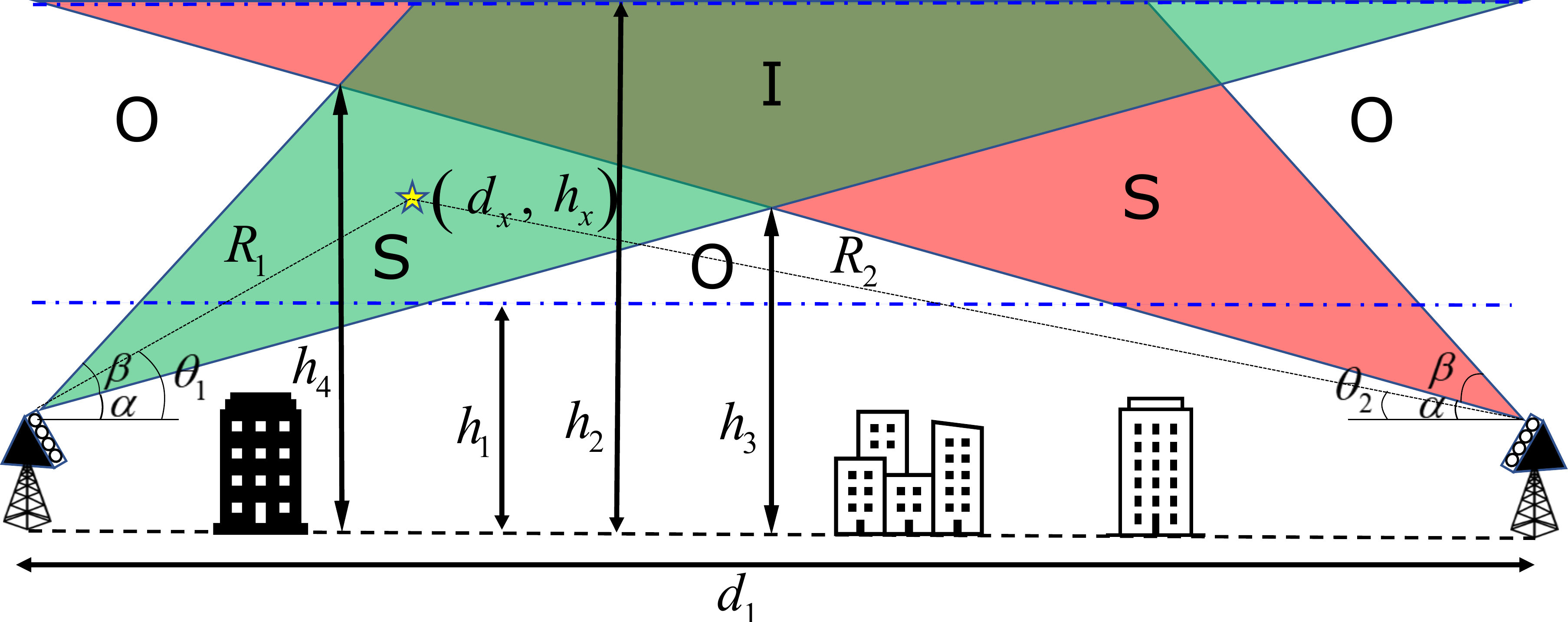}\label{fig:illu_c3}}\vspace{-1mm}
    
    \subfloat[Case 4]{\includegraphics[width=0.47\textwidth]{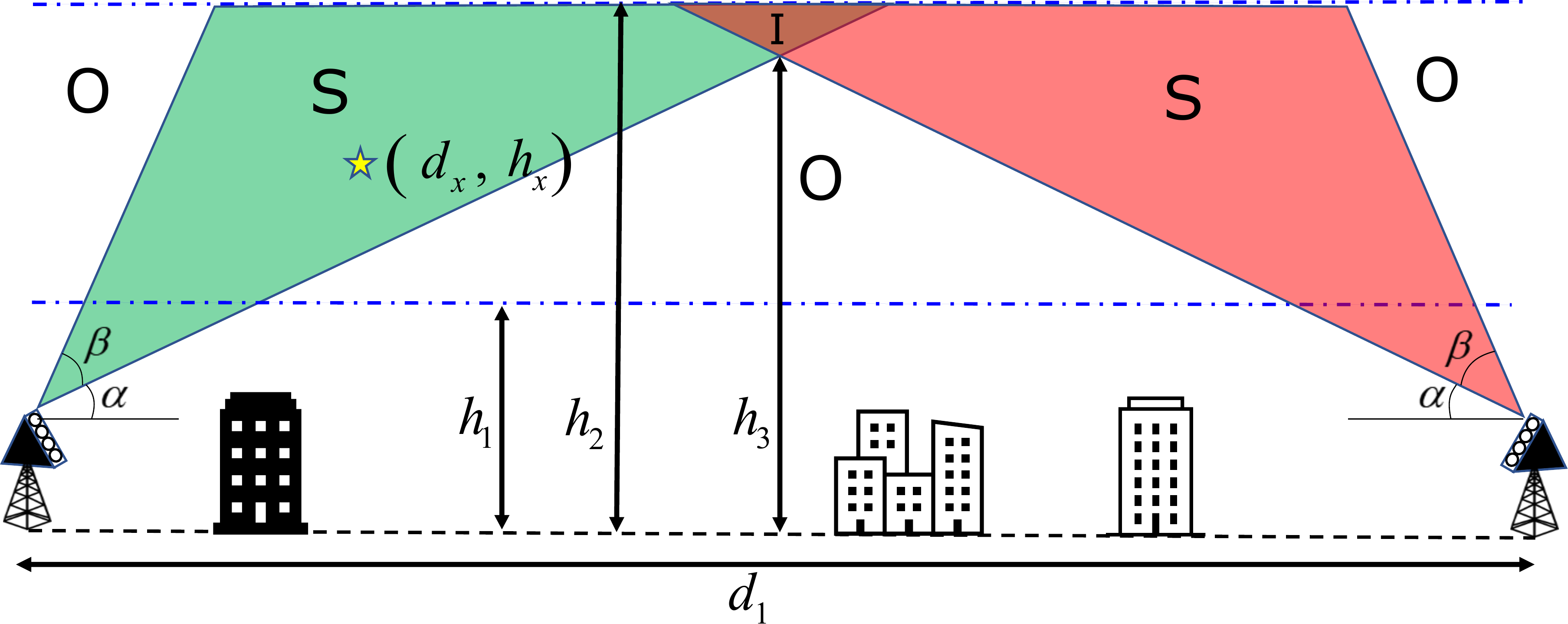}}\vspace{-1mm}
    
    \subfloat[Case 5]{\includegraphics[width=0.47\textwidth]{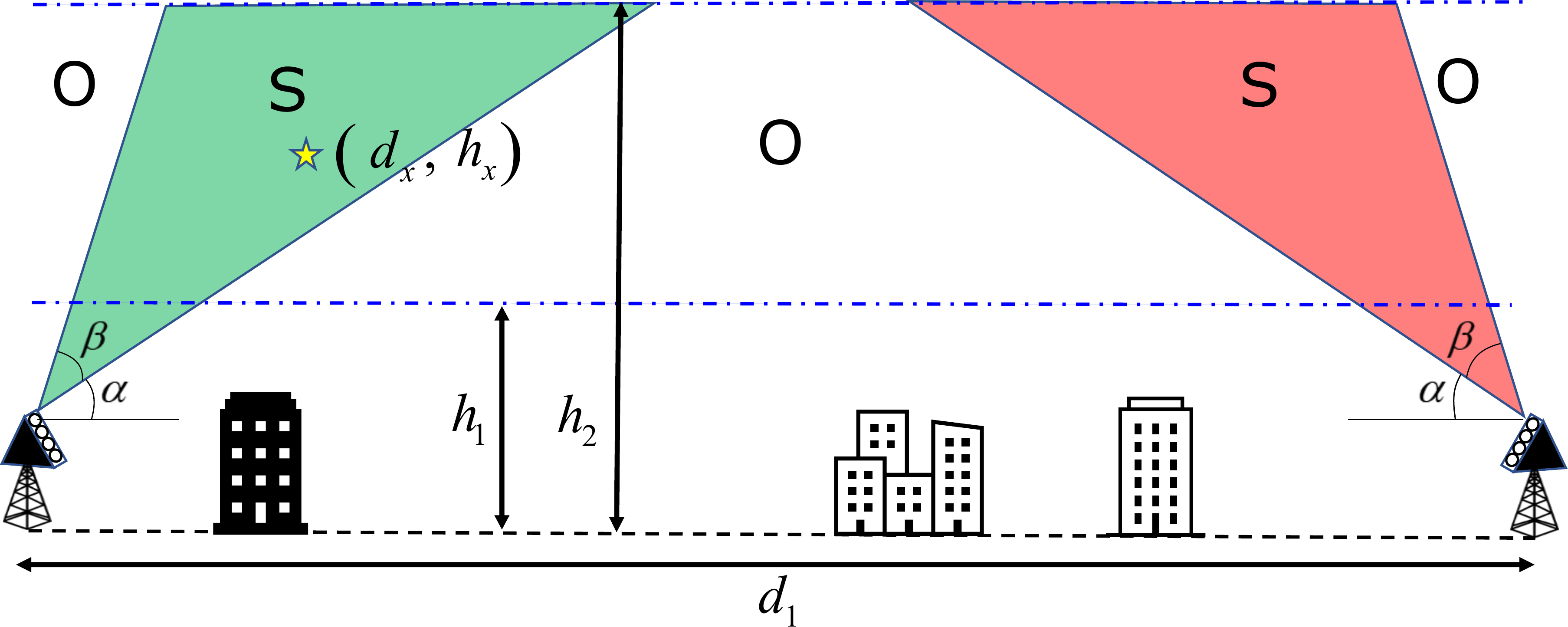}}
    \caption{The 2D coordinate drone corridor bounded by $h_1$ and $h_2$, and the  five different corridor coverage scenarios as the uptilt angle $\alpha$ grows.}
    \label{fig:illu}\vspace{-5mm}
\end{figure}

We consider a 2D coordinate system for simplicity where the drone corridor is covered by two base stations (BS) as shown in Fig.~\ref{fig:illu}. The horizontal distance between the two BSs is $d_1$, and the lowest and the highest altitudes of the drone corridor are $h_1$ and $h_2$, respectively. The BSs are equipped with a vertically directional antenna. The center angle of the directional beam is controlled by the uptilt angle $\alpha$ and the beamwidth of the antenna pattern is given by $\beta$. The height of the BSs is assumed to 0~[m] without loss of generality.

\subsection{Probability Density Functions of the Location of UAVs}

We assume that C-UAVs are uniformly distributed in the drone corridor area and served by the nearest BS. Then, the probability density function (PDF) of the horizontal ($d_{\rm x}$) and vertical distance ($h_{\rm x}$) of the UAVs to their serving BSs are given by
\begin{align}
    f_{d_{\rm x}}(d_{\rm x})&=\frac{2}{d_1},\;[0<d_{\rm x}<d_1/2],\\
    f_{h_{\rm x}}(h_{\rm x})&=\frac{1}{h_2-h_1},\;[h_1<h_{\rm x}<h_2].
\end{align}
The elevation angles from the serving BS and the neighboring BS can be expressed as
\begin{align}\label{eq:theta}
    \theta_1=\tan^{-1}\left(\frac{h_{\rm x}}{d_{\rm x}}\right),\;\theta_2=\tan^{-1}\left(\frac{h_{\rm x}}{d_1-d_{\rm x}}\right).
\end{align}
Then, the PDF of random variable $\theta_1$ given $h_{\rm x}$ can be derived by the PDF transformation function as
\begin{align}
    f_{\theta_1}(\theta_1|h_{\rm x})&\overset{(a)}{=}\sum_{\theta_1=\tan^{-1}\left(\frac{h_{\rm x}}{d_{\rm x}}\right)} f_{d_{\rm x}}\left(\frac{h_{\rm x}}{\tan\theta_1}\right)\left|\frac{\partial d_{\rm x}}{\partial\theta_1}\right|\nonumber\\
    &\overset{(b)}{=}f_{d_{\rm x}}\left(\frac{h_{\rm x}}{\tan\theta_1}\right)|h_{\rm x}\cot^2\theta_1+h_{\rm x}|\nonumber\\
    &=\frac{2h_{\rm x}}{d_{1}}\csc^2\theta_1,\;\left[\tan^{-1}\left(\frac{2h_{\rm x}}{d_{1}}\right)<\theta_1<\frac{\pi}{2}\right],
\end{align}
where (a) comes from $d_{\rm x}=\frac{h_{\rm x}}{\tan\theta_1}$ and (b) comes from the fact that there is only one $d_{\rm x}$  satisfying $\theta_1=\tan^{-1}\left(\frac{h_{\rm x}}{d_{\rm x}}\right)$.

\subsection{Rectangular Beampattern Model}

We adopt the rectangular shape beam model for the directional antenna pattern, which is given by~~\cite{kovalchukov2018analyzing,yi2019unified}
\begin{align}
    g_{\rm y}(\theta_{\rm y})&=
    \begin{cases}
    G & \text{if }\alpha<\theta_{\rm y}<\alpha+\beta\\
    0 & \text{otherwise}
    \end{cases}~,\\
    \alpha&>0,\; \alpha+\beta<\frac{\pi}{2}~,\label{eq:bp_const}
\end{align}
where $g_{\rm y}(\theta_{\rm y})$ and $G$ denote the antenna pattern of the BS and the maximum antenna gain, respectively, and \eqref{eq:bp_const} indicates that the main beam is not steered downward and does not go beyond 90 degrees. In addition, ${\rm y}\in\{1,2\}$, and ${\rm y}=1$, ${\rm y}=2$ indicate the serving BS and the neighboring BS, respectively. We assume that the two BSs use the same uptilt angle ($\alpha$) and beamwidth $\beta$. This model is simple but mathematically tractable and reflects adequately the effect of the directional beams.

\subsection{Case Study for Increasing Uptilt Angle}

We divide the drone corridor area into three regions depending on which beams cover portions of the drone corridor. As shown in Fig.~\ref{fig:illu}, the area served only by the green or red beam of the serving BS is the `beam served region without interference' (S), and the overlapping area of two beams is the `interference region' (I), and the rest area (including that in the beam of the neighboring BS) is the 'beam outage region' (O). The illustrations of case 1 to case 5 show how the three areas change as the uptilt angle $\alpha$ increases. We found that it is necessary to analyze the performance case by case depending on the value of the uptilt angle. For example, the drone corridor area can be divided into the `beam outage region' (O) and the `interference region' (I) in case 1, while the `beam served region without interference' (S) appears from $h_1$ to $h_4$ in case 2. Then, the `beam outage region' (O) on cell edge area appears from $h_1$ to $h_3$ in case 3.
As shown in  Fig.~\ref{fig:illu_c3}, we denote the heights of the crossing points of the beams from the two BSs as $h_3$ (center) and $h_4$ (side), given by:
\begin{align}
     h_3&=\frac{d_1}{2}\tan(\alpha),\;  h_{4}=\frac{d_1}{\cot(\alpha)+\cot(\alpha+\beta)}~.
\end{align}
The 3D distances between a UAV and the serving BS are denoted by $R_1$, while the distance between a UAV and the neighboring BS is given by $R_2$:
\begin{align}
     R_1&=\frac{h_{\rm x}}{\sin(\theta_1)},\;R_2=\frac{h_{\rm x}}{\sin(\theta_2)}~.
\end{align}
We consider free-space pathloss model, which is expressed as $\mathsf{PL}=\left(\frac{4\pi R_1}{\lambda}\right)^2$. Then, signal-to-interference-plus-noise ratio (SINR) can be written as
\begin{align}\label{eq:SINR}
    \mathsf{SINR}&=\frac{kg_1(\theta_1)/(R_1)^2}{kg_2(\theta_2)/(R_2)^2+N_0},\;k=\frac{\mathsf{P}_{\mathsf{Tx}}\lambda^2}{16\pi^2},
\end{align}
where $\mathsf{P}_{\mathsf{Tx}}$, $\lambda$, $N_0$ denote transmit power, wave-length, and noise power respectively, and $g_1(\theta_1)$, $g_2(\theta_2)$ indicate antenna gain of the serving and the neighboring BSs.

\section{SINR Outage Probability Analysis}

In this section, we derive the closed-form expression of SINR outage probability as a function of the uptilt angle ($\alpha$). We can find the optimal uptilt angle that minimizes the outage probability from the obtained results. The SINR outage probability can be defined as follows
\begin{align}\label{eq:p_out_def}
    \mathsf{Pr}_{\mathsf{out}}& = 1-\mathsf{Pr}_{\mathsf{in}}=\mathsf{Pr}(\mathsf{SINR}<\tau)~,
\end{align}
where $\tau$ is the SINR threshold. We derive the SINR outage probability by obtaining the probability that a UAV is located at the beam outage region. Since the analytical expression can be different depending on the uptilt angle as shown in Fig.~\ref{fig:illu}, we derive the analytical expressions separately considering three distinct cases. Note that we combine Case 1 and Case 2, and Case 3 and Case 4 as a two distinct cases.

\subsection{Case 1 $\&$ 2 $(h_1 > h_3)$}

In case 1 (Fig.~\ref{fig:illu_c1}) and 2 (Fig.~\ref{fig:illu_c2}), the green and red beams covers cell-edge drones ($d_{\rm x}=\frac{d_1}{2}$) for all heights from $h_1$ to $h_2$. In these cases, the beam served region with or without interference  starts from the UAV at the center $\left(d_{\rm x}= \frac{d_1}{2},~  \theta_1=\tan^{-1}\left(\frac{2h_{\rm x}}{d_{1}}\right)\right)$ and ends to the UAV at the left main beam edge ($\theta_1=\alpha+\beta$) for all heights. We assume that the threshold level of SINR ($\tau$) is designed to a moderate level, so that UAVs in the interference region are satisfied with SINR criteria. The probability that a UAV is located at the beam served region can be expressed as
\begin{align}
    \mathsf{Pr}_{\rm in}&=\int_{h_1}^{h_2}\int_{\tan^{-1}\left(\frac{2h_{\rm x}}{d_{1}}\right)}^{\alpha+\beta} f_{\theta_1}(\theta_1|h_{\rm x})f_{h_{\rm x}}(h_{\rm x})\partial\theta_1\partial h_{\rm x}\nonumber\\
    &=\frac{-(h_1+h_2)}{d_1}\cot(\alpha+\beta)+1.
\end{align}
Then, SINR outage probability can be written as
\begin{align}\label{eq:p_out_c1}
    \mathsf{Pr}_{\rm out}&=1-\mathsf{Pr}_{\rm in}=\frac{(h_1+h_2)}{d_1}\cot(\alpha+\beta).
\end{align}
We can also obtain the first-order partial derivative with respect to $\alpha$ as follows
\begin{align}
    \frac{\partial\mathsf{Pr}_{\rm out}(\alpha)}{\partial\alpha}&=\frac{-(h_1+h_2)}{d_1}\csc^2(\alpha+\beta).
\end{align}

\subsection{Case 3 $\&$ 4 $(h_1 < h_3 < h_2)$}

In case 3 and 4, the elevation angle range of the beam served region can be divided into two parts; from $h_1$ to $h_3$ and from $h_3$ to $h_2$. The  SINR outage probability can be expressed as

\begin{align}\label{eq:p_out_c2}
    \mathsf{Pr}_{\rm out}&=1-\int_{h_1}^{h_3}\int_{\alpha}^{\alpha+\beta}f_{\theta_1}(\theta_1|h_{\rm x})f_{h_{\rm x}}(h_{\rm x})\partial\theta_1\partial h_{\rm x}\nonumber\\
    &-\int_{h_3}^{h_2}\int_{\tan^{-1}\left(\frac{2h_{\rm x}}{d_{1}}\right)}^{\alpha+\beta} f_{\theta_1}(\theta_1|h_{\rm x})f_{h_{\rm x}}(h_{\rm x})\partial\theta_1\partial h_{\rm x}\nonumber\\
    &=1+\frac{(h_1+h_2)}{d_1}\cot(\alpha+\beta)\nonumber\\
    &-\frac{\frac{d_1^2}{4}\tan^2(\alpha)-h_1^2}{d_1(h_2-h_1)}\cot(\alpha)-\frac{h_2-\frac{d_1}{2}\tan(\alpha)}{h_2-h_1}.
\end{align}
We can also obtain the first-order partial derivative with respect to $\alpha$ as follows
\begin{align}
    \frac{\partial\mathsf{Pr}_{\rm out}(\alpha)}{\partial\alpha}&=-\frac{(h_1+h_2)}{d_1}\csc^2(\alpha+\beta)\nonumber\\
    &+\frac{d_1}{4(h_2-h_1)}\sec^2(\alpha)-\frac{(h_1)^2}{d_1(h_2-h_1)}\csc^2(\alpha)~.
\end{align}

\subsection{Case 5 $(h_3 > h_2)$}
In case 5, the elevation angle range of the beam served area is from $\alpha$ to $\alpha+\beta$ for all heights. The SINR outage probability can be written as
\begin{align}\label{eq:p_out_c3}
    \mathsf{Pr}_{\rm out}&=1-\int_{h_1}^{h_2}\int_{\alpha}^{\alpha+\beta}f_{\theta_1}(\theta_1|h_{\rm x})f_{h_{\rm x}}(h_{\rm x})\partial\theta_1\partial h_{\rm x}\nonumber\\
    &=1+\frac{(h_1+h_2)}{d_1}\left(\cot(\alpha+\beta)-\cot(\alpha)\right).
\end{align}
The first-order partial derivative is given by
\begin{align}
    \frac{\partial\mathsf{Pr}_{\rm out}(\alpha)}{\partial\alpha}&=-\frac{(h_1+h_2)}{d_1}\left(\csc^2(\alpha+\beta)-\csc^2(\alpha)\right).
\end{align}

Now that we obtain the SINR outage probability ($\mathsf{Pr}_{\rm out}$) and the first-order partial derivative, we can find the optimal uptilt angle ($\alpha$) by using an optimization method like gradient descent if the function is convex.

\section{Average SINR Analysis}

In this section, we calculate the average SINR of UAVs in the drone corridor. We separately derive the expressions depending on the cases in Fig.~\ref{fig:illu} as the uptilt angle increases. From \eqref{eq:SINR}, the definition of the average SINR of drones in the whole drone corridor area is given by
\begin{align}
    \mathsf{SINR}_{\rm avg}&=\mathbb{E}\left[\mathsf{SINR}\right]\label{eq:SINR_avg_def}\\
    &=\int_{h_1}^{h_2}\int_{\tan^{-1}\left(\frac{2h_{\rm x}}{d_{1}}\right)}^{\frac{\pi}{2}}\left(\frac{kg_1(\theta_1)/(R_1)^2}{kg_2(\theta_2)/(R_2)^2+N_0}\right)\nonumber\\
    &\times f_{\theta_1}(\theta_1|h_{\rm x})f_{h_{\rm x}}(h_{\rm x})\partial\theta_1\partial h_{\rm x}.
\end{align}

\subsubsection{Case 1 $(h_1 > h_3\;\&\;h_1>h_4)$}
The whole beam served area is 'interference region' in case 1. The average SINR can be written as

\begin{align}\label{eq:SINR_avg_c1}
    &\mathsf{SINR}_{\rm avg}\nonumber\\
    &=\int_{h_1}^{h_2}\int_{\tan^{-1}\left(\frac{2h_{\rm x}}{d_{1}}\right)}^{\alpha+\beta}\left(\frac{(R_2)^2}{(R_1)^2}\right)f_{\theta_1}(\theta_1|h_{\rm x})f_{h_{\rm x}}(h_{\rm x})\partial\theta_1\partial h_{\rm x}\nonumber\\
    &=\int_{h_1}^{h_2}\left(\frac{2d_1}{h_{\rm x}(h_2-h_1)}\right)\left(\alpha+\beta-\tan^{-1}\left(\frac{2h_{\rm x}}{d_{1}}\right)\right)\nonumber
\end{align}
\begin{align}
    &-\left(\frac{4}{h_2-h_1}\right)\nonumber\\
    &\times\left(\log\left(\sin(\alpha+\beta)\right)-\log\left(\sin\left(\tan^{-1}\left(\frac{2h_{\rm x}}{d_{1}}\right)\right)\right)\right)\nonumber\\
    &-\left(\frac{2h_{\rm x}}{d_1(h_2-h_1)}\right)\left(\cot(\alpha+\beta)-\frac{d_{1}}{2h_{\rm x}}\right)\partial h_{\rm x}.
\end{align}
Note that we assume the interference dominant condition so that we neglect the noise term. We neglect the `beam outage region' where $\theta_1$ is from $\alpha+\beta$ to $\frac{\pi}{2}$ in the calculation, since $\mathsf{SINR}\approx 0$.

\subsubsection{Case 2 $(h_3 < h_1 < h_4)$}

From case 2 to case 5, we only consider `beam served region without interference' (green region) since the average SINR is dominant by that region. Then, the average SINR can be expressed as
\begin{align}\label{eq:SINR_avg_c2}
    \mathsf{SINR}_{\rm avg}&=\int_{h_1}^{h_4}\int_{\gamma}^{\alpha+\beta}\left(\frac{kG}{N_0(R_1)^2}\right)f_{\theta_1}(\theta_1|h_{\rm x})f_{h_{\rm x}}(h_{\rm x})\partial\theta_1\partial h_{\rm x}\nonumber\\
    &=\int_{h_1}^{h_4}\left(\frac{2kG}{d_1N_0(h_2-h_1)h_{\rm x}}\right)\nonumber\\
    &\times\left(\alpha+\beta-\cot^{-1}\left(\frac{d_1}{h_{\rm x}}-\cot(\alpha)\right)\right)\partial h_{\rm x},
\end{align}
where $\gamma=\cot^{-1}\left(\frac{d_1}{h_{\rm x}}-\cot(\alpha)\right)$ indicates the elevation angle at the right end point of green region. We can derive $\gamma$ as follow: the relation between $\theta_1$ and $\theta_2$ is written from \eqref{eq:theta} as
\begin{align}
    \cot(\theta_2)&=\frac{d_1}{h_{\rm x}}-\cot(\theta_1).
\end{align}
Then, we can obtain $\gamma$ by plugging in $\theta_1=\gamma$, $\theta_2=\alpha$.

\subsubsection{Case 3 $(h_1 < h_3<h_2\;\&\; h_2 > h_4)$}

The elevation angle range that we are concerned is different from $h_1$ to $h_3$ and from $h_3$ to $h_4$. Then, the average SINR can be written as
\begin{align}\label{eq:SINR_avg_c3}
    &\mathsf{SINR}_{\rm avg}
    =\int_{h_1}^{h_3}\int_{\alpha}^{\alpha+\beta}\left(\frac{kG}{N_0(R_1)^2}\right)f_{\theta_1}(\theta_1|h_{\rm x})f_{h_{\rm x}}(h_{\rm x})\partial\theta_1\partial h_{\rm x}\nonumber\\
    &+\int_{h_3}^{h_4}\int_{\gamma}^{\alpha+\beta}\left(\frac{kG}{N_0(R_1)^2}\right)f_{\theta_1}(\theta_1|h_{\rm x})f_{h_{\rm x}}(h_{\rm x})\partial\theta_1\partial h_{\rm x}\nonumber\\
    &=\int_{h_1}^{h_3}\left(\frac{2kG\beta}{d_1N_0(h_2-h_1)h_{\rm x}}\right)\partial h_{\rm x}+\int_{h_3}^{h_4}\left(\frac{2kG}{d_1N_0(h_2-h_1)h_{\rm x}}\right)\nonumber\\
    &\times\left(\alpha+\beta-\cot^{-1}\left(\frac{d_1}{h_{\rm x}}-\cot(\alpha)\right)\right)\partial h_{\rm x}.
\end{align}

\subsubsection{Case 4 $(h_1 < h_3<h_2\;\&\; h_2 < h_4)$}

The average SINR can be written as
\begin{align}\label{eq:SINR_avg_c4}
    &\mathsf{SINR}_{\rm avg}
    =\int_{h_1}^{h_3}\int_{\alpha}^{\alpha+\beta}\left(\frac{kG}{N_0(R_1)^2}\right)f_{\theta_1}(\theta_1|h_{\rm x})f_{h_{\rm x}}(h_{\rm x})\partial\theta_1\partial h_{\rm x}\nonumber\\
    &+\int_{h_3}^{h_2}\int_{\gamma}^{\alpha+\beta}\left(\frac{kG}{N_0(R_1)^2}\right)f_{\theta_1}(\theta_1|h_{\rm x})f_{h_{\rm x}}(h_{\rm x})\partial\theta_1\partial h_{\rm x}\nonumber\\
    &=\int_{h_1}^{h_3}\left(\frac{2kG\beta}{d_1N_0(h_2-h_1)h_{\rm x}}\right)\partial h_{\rm x}+\int_{h_3}^{h_2}\left(\frac{2kG}{d_1N_0(h_2-h_1)h_{\rm x}}\right)\nonumber\\
    &\left(\alpha+\beta-\cot^{-1}\left(\frac{d_1}{h_{\rm x}}-\cot(\alpha)\right)\right)\partial h_{\rm x}.
\end{align}

\subsubsection{Case 5 $(h_3 > h_2)$}

The average SINR can be written as
\begin{align}\label{eq:SINR_avg_c5}
    \mathsf{SINR}_{\rm avg}
    &=\int_{h_1}^{h_2}\int_{\alpha}^{\alpha+\beta}\left(\frac{kG}{N_0(R_1)^2}\right)f_{\theta_1}(\theta_1|h_{\rm x})f_{h_{\rm x}}(h_{\rm x})\partial\theta_1\partial h_{\rm x}\nonumber\\
    &=\int_{h_1}^{h_2}\left(\frac{2kG\beta}{d_1N_0(h_2-h_1)h_{\rm x}}\right)\partial h_{\rm x}.
\end{align}

\section{Numerical Results} \label{sec:results}

\begin{table}[!t]
\renewcommand{\arraystretch}{1.1}
\caption{Simulation settings}
\label{table:settings}
\centering
\begin{tabular}{lc}
\hline
Parameter & Value \\
\hline\hline
Transmit power ($\mathsf{P}_\mathsf{Tx}$) & $30$~dBm\\
Horizontal distance between BSs ($d_1$) & $1000$~m \\
Minimum drone corridor height ($h_1$) & $100$~m \\
Maximum drone corridor height ($h_2$) & [$200$, $300$, $400$, $500$]~m \\
Threshold level of SINR ($\tau$) & $-3$~dB \\
Maximum antenna gain ($G$) & $297.6$/$\beta$~dB~\cite{venugopal2016device} \\
Carrier frequency & $3$~GHz \\
Bandwidth (BW) & $100$~MHz\\
Thermal noise (TN) & $-174$~dBm/Hz\\
Noise figure (NF) & $9$~dB\\\hline
\hline
\end{tabular}
\vspace{-0.15in}
\end{table}

In this section, we present simulation results to verify above analysis and evaluate the performance depending on the uptilt angle. The key parameters are listed in Table~I. In general, the maximum antenna gain ($G$) is inversely proportional to the beamwidth ($\beta$). In other words, as the beam becomes sharper, the maximum beam gain is higher. We adopt the result in \cite[Table II]{venugopal2016device} to calculate the antenna gain. Noise power is calculated by $N_0=(\text{TN})+10\log_{10}(\text{BW})+(\text{NF})~\text{[dBm]}$.

\begin{figure}[t]
	\centering
	\vspace{-0.0in}
	\subfloat[SINR outage probability with $h_2 = 300$~m.]{\includegraphics[width=0.48\textwidth]{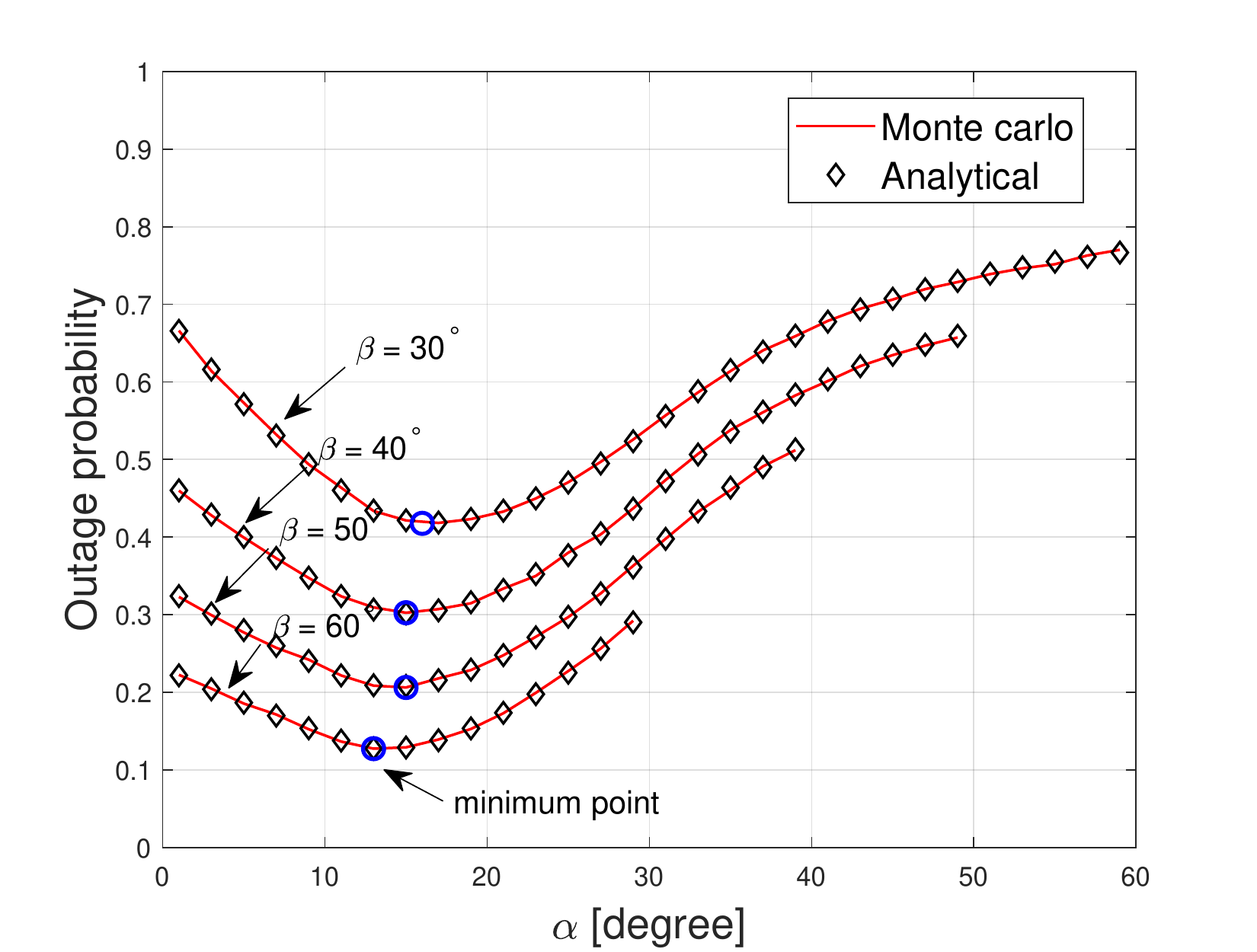}}\label{fig:outage_sim(a)}
	\subfloat[SINR outage probability with $\beta = 50^{\circ}$.]{\includegraphics[width=0.48\textwidth]{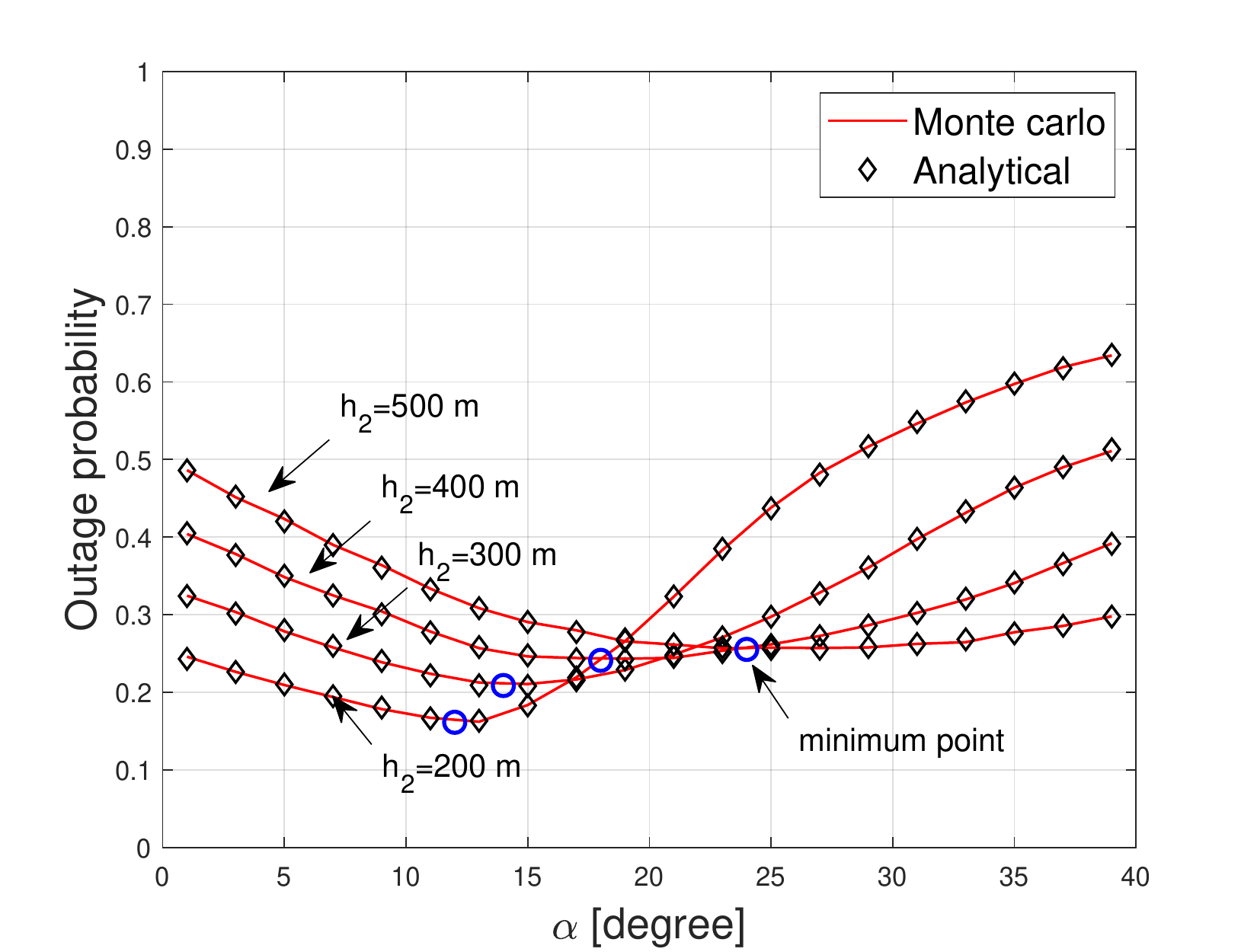}}\label{fig:outage_sim(b)}
	\caption{SINR outage probability depending on the uptilt angle ($\alpha$).}\label{fig:outage_sim}\vspace{-2mm}
\end{figure}

\begin{figure}[t]
	\centering
	\vspace{-0.0in}
	\subfloat[Average SINR with $h_2 = 300$~m.]{\includegraphics[width=0.48\textwidth]{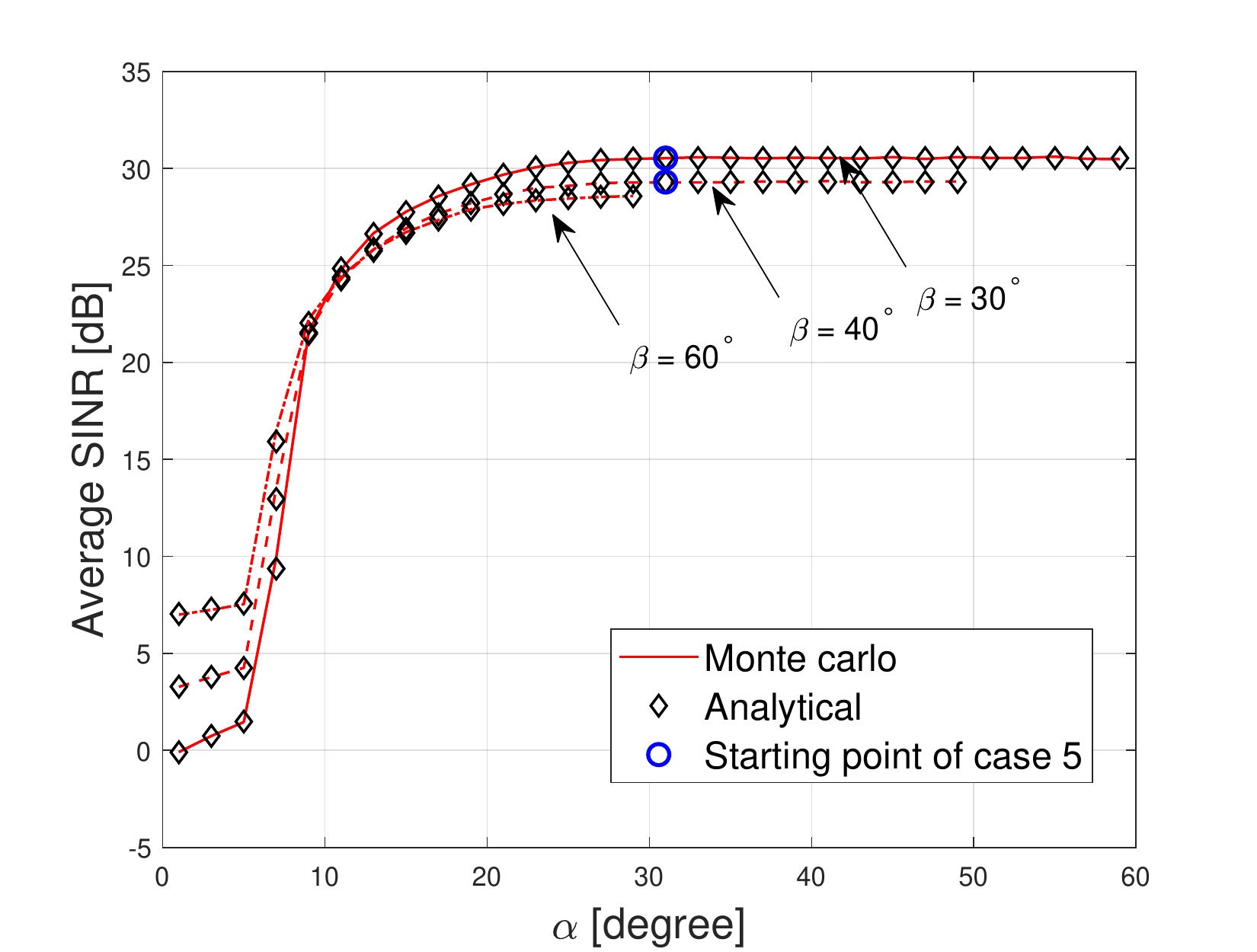}}\vspace{-3mm}
	
	\subfloat[Average SINR with $\beta = 50^{\circ}$.]{\includegraphics[width=0.48\textwidth]{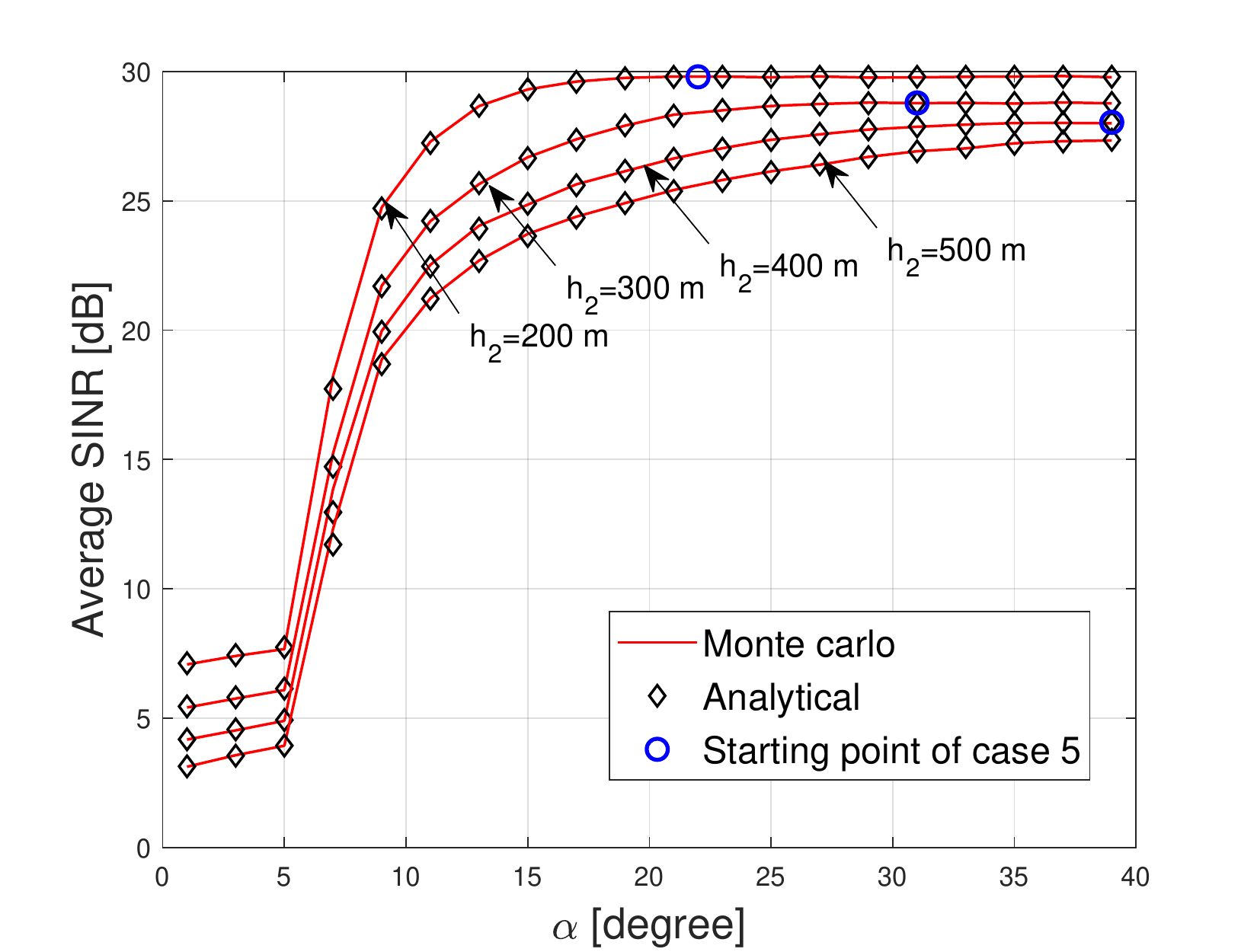}}
	\caption{Average of SINR depending on the uptilt angle ($\alpha$).}\label{fig:SINR_avg}\vspace{-2mm}
\end{figure}

Fig.~\ref{fig:outage_sim} shows the SINR outage probability with respect to the uptilt angle. We use \eqref{eq:p_out_def} to obtain Monte Carlo results and use \eqref{eq:p_out_c1}, \eqref{eq:p_out_c2}, \eqref{eq:p_out_c3} to get analytical results.
It is observed that the SINR outage probability is a convex function with respect to $\alpha$, so that we can obtain the global minimum by the first-order partial derivative, and the optimal point that minimizes the outage probability is different depending on beamwidth ($\beta$) and the maximum drone corridor height ($h_2$). We also observe that the performance with the optimal uptilt angle ($\alpha$) improves as the beamwidth increases and the maximum drone corridor height decreases. 

In Fig.~\ref{fig:SINR_avg}, it is shown that the average SINR increases as the uptilt angle grows and converges to the maximum value. The Monte Carlo results are obtained by \eqref{eq:SINR_avg_def} and the analytical results are obtained by \eqref{eq:SINR_avg_c1}-\eqref{eq:SINR_avg_c5}. Intuitively, the best average SINR is achieved as Case 5 emerges, where the whole beamwidth (from $\theta_1=\alpha$ to $\theta_1=\alpha+\beta$) is inside of the drone corridor without interference for all heights from $h_1$ to $h_2$. It is also observed that the maximum average SINR increases as beamwidth ($\beta$) and the maximum height of the drone corridor ($h_2$) decreases.

\section{Conclusion}\label{sec:conclusion}
In this paper, we study the optimal uptilt angle of the antenna for cellular-connected drone corridor communication. We consider two BSs with 2D coordinates and the directional antenna pattern. We derive the closed-form expressions of SINR outage probability and average SINR to evaluate the performance. We found that the SINR outage probability is a convex function of the uptilt angle, and the average SINR is maximized from a certain degree of the uptilt angle which is the starting point of Case 5 in our five identified scenarios. Furthermore, it is shown that wide beamwidth is preferred in terms of the outage probability, while narrow beamwidth is desired from the perspective of the average SINR. A lower maximum height of the drone corridor will benefit both performance criteria. Although our study is limited to the 2D coordinate system, considering a scenario where the BSs are deployed along the drone corridor, our results and findings can be conveniently extended to 3D drone corridor design.   

\bibliographystyle{IEEEtran} 
\bibliography{IEEEabrv,bibfile}
  
\end{document}